\documentclass[twocolumn,11pt]{mwart}
\usepackage{graphicx} 
\usepackage[margin=20mm, top=25mm, bottom=25mm]{geometry}
\usepackage{xcolor}
\usepackage{antpolt} 
\usepackage{feynmf} 
\usepackage{caption} 
\usepackage{polski}
\usepackage[utf8]{inputenc}

\usepackage{icomma} 

\usepackage[
backend=biber,
style=phys,
sorting=none
]{biblatex}
\usepackage{amsmath}
\usepackage{amssymb}
\usepackage{tikz}
\usepackage{gnuplottex}

\usepackage{xspace}

\title{Bilard protonowy w LHC}
\author{\textit{Rafał Staszewski}\\[1em]
Instytut Fizyki Jądrowej PAN}
\date{\today}

\makeatletter
\renewcommand{\@maketitle}{%
  \newpage
  \null
  \vskip 2em%
  \begin{center}%
    {\huge\bfseries \@title \par}%
    \vskip 2em%
    {\Large
      \lineskip .5em%
      \begin{tabular}[t]{c}%
        \@author
      \end{tabular}\par}%
    \vskip 1.5em%
    {\Large \@date \par}%
  \end{center}%
  \par
  \vskip 2.5em}
\makeatother

\begin{document}

\maketitle

\begin{abstract}
Proton--proton elastic scattering at high energy is a process with simple kinematics but surprisingly complex dynamics.
Measurements of these processes at particle colliders require dedicated detectors and unusual measurement conditions.

In the article, I describe the physics of elastic processes and the experimental method in the context of the measurement performed by the ATLAS Collaboration at the LHC accelerator.
I present the key elements of the data analysis and discuss the  consequences of the obtained results for our understanding of the strong interactions.

\noindent\hfill\rule{3cm}{0.5pt}\hfill\hfill

\smallskip
Rozpraszanie elastyczne proton--proton przy wysokich energiach jest procesem o prostej kinematyce, lecz zaskakująco złożonej dynamice.
Pomiary tych procesów na zderzaczach cząstek wymagają specjalnie zaprojektowanych detektorów oraz nietypowych warunków pomiarowych.

W artykule opisuję fizykę procesów elastycznych oraz metodę eksperymentalną pomiaru wykonanego w ramach Współpracy ATLAS na akceleratorze LHC. 
Przedstawiam kluczowe elementy analizy danych, a także omawiam konsekwencje uzyskanego wyniku dla naszego rozumienia oddziaływań silnych.
\end{abstract}

\section{Rozpraszanie elastyczne}
W bilard grał prawie każdy. 
Uderzając kijem w~białą bilę mamy za zadanie wbić do łuz bile innych kolorów.
Zderzenia kul bilardowych są klasycznym przykładem zderzeń sprężystych, czyli takich, w których zachowany jest nie tylko całkowity pęd, ale również całkowita energia kinetyczna. 
Zderzenia sprężyste występują również w~świecie kwantowym,
choć w tym kontekście częściej używa się nazwy \emph{rozpraszanie elastyczne}.
Pomiar własności takich procesów jest jedną z~podstawowych metod badania cząstek elementarnych i~ich oddziaływań.
Wykorzystuje się ją w badaniach prowadzonych na akceleratorze LHC w Europejskiej Organizacji Badań Jądrowych CERN w Szwajcarii.
Akcelerator ten, choć uruchomiony przede wszystkich z myślą o odkrywaniu nowych cząstek, umożliwia również pomiary procesów rozpraszania elastycznego przy najwyższych energiach dostępnych w warunkach laboratoryjnych.
W badania takie, prowadzone w ramach eksperymentu ATLAS \cite{ATLAS}, zaangażowana jest grupa z~Instytutu Fizyki Jądrowej PAN w Krakowie.
Niniejszy artykuł opisuje pomiar rozpraszania elastycznego w zderzeniach proton--proton przy sumarycznej energii w układzie środka masy równej 13 TeV \cite{STDM-2018-08}.

\begin{figure}
\centering

\includegraphics{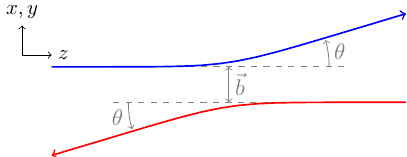}

\vspace{3ex}

\begin{tikzpicture}
   \draw[->] (-3.5, -1.2) -- +(0.5, 0) node[right] {\small $x$};
   \draw[->] (-3.5, -1.2) -- +(0, 0.5) node[above] {\small $y$};
   \draw[fill=red!40] (-2,  0.5) circle (0.7);
   \draw[fill=blue!40] (+2, -0.5) circle (0.7);
   \draw[gray,<->] (-2, 0.5) -- node[above] {$\vec{b}$} (+2, -0.5);
\end{tikzpicture}
\caption{Definicja podstawowych wielkości charakteryzujących rozpraszanie elastyczne: parametru zderzenia i kąta rozproszenia.}
\label{fig:impactpar}
\end{figure}

Kinematyka oddziaływania elastycznego jest bardzo prosta w układzie środka masy --
pędy cząstek zmieniają kierunek, ale nie zmieniają swoich wartości (długości wektora).
Kąt $\theta$ pomiędzy prędkością początkową i końcową nazywany jest kątem rozproszenia.
Ważną wielkością jest parametr zderzenia $b$, który definiuje się jako odległość poprzeczną pomiędzy początkowymi torami cząstek.
Definicje tych wielkości pokazane zostały na rysunku \ref{fig:impactpar}.

W przeciwieństwie do zderzeń kul na stole bilardowym, elastyczne zderzenia protonów, $pp\to pp$, zachodzą w trzech wymiarach.
Aby w pełni scharakteryzować konkretny przypadek takiego oddziaływania potrzebne są dwa kąty: nie tylko polarny kąt rozproszenia $\theta$ ale i kąt azymutalny $\varphi$.
W~eksperymentach przy najwyższej energii, na przykład na akceleratorze LHC, ale i na wielu wcześniejszych akceleratorach, wiązki protonowe często nie są spolaryzowane.
Wtedy rozkład kąta $\varphi$, jako zmiennej losowej, jest jednorodny i~nie daje on żadnej informacji o samym oddziaływaniu.
Cała informacja o dynamice procesu jest zawarta w rozkładzie kąta $\theta$.
Ponieważ wartość kąta rozproszenia jest różna w różnych układach odniesienia, w praktyce zamiast niego używa się tak zwanej zmiennej $t$ Mandelstama, zdefiniowanej jako kwadrat zmiany czteropędu cząstki podczas rozproszenia:
\begin{equation}
t = (P_k - P_p)^2,
\end{equation}
gdzie $P_p$ jest czteropędem początkowym, a $P_k$ czteropędem końcowym jednego z protonów (obojętnie którego).
Zmienna ta jest kwadratem czterowektora, a więc niezmiennikiem przekształceń Lorentza.

W układzie środka masy wartość zmiennej $t$ wyraża się wzorem:
\begin{equation}
\label{eq:t}
t = - 2 p^2 (1 - \cos\theta),
\end{equation}
gdzie $p$ jest wartością pędu cząstki (w tym układzie odniesienia wartość ta jest równa dla obu cząstek i nie zmienia się podczas oddziaływania).
Dla małych kątów, a~takie są najczęściej badane, można zastosować przybliżenie $t = - p^2 \theta^2$.

Warto zwrócić uwagę, ze wartość zmiennej $t$ jest zawsze mniejsza lub równa zero.
Dlatego często dla wygody przedstawia się wykresy nie w funkcji $t$, ale $|t|$ lub  $-t$.
Równie często używa sie zwrotu ,,małe $t$'' lub ,,duże $t$'',  mając na myśli wartości bezwzględne: ,,małe $|t|$'' i ,, duże $|t|$''.

\newcommand{\vp}{\ensuremath{{\vec p}_T}\xspace}
\newcommand{\vb}{\ensuremath{\vec b}\xspace}
Oprócz prostej i Lorentzowsko niezmienniczej definicji, ważną zaletą zmiennej $t$ jest jej bliski związek z pędem poprzecznym, czyli rzutem pędu na płaszczyznę prostopadłą do osi zderzenia.
W fizyce cząstek standardową konwencją jest wybór układu współrzędnych o osi $z$ wzdłuż kierunku zderzenia, 
a~osiami $x$ i $y$ w płaszczyźnie poprzecznej, tak jak  zostało to pokazane na rysunku \ref{fig:impactpar}.
Wtedy:
\begin{equation}
    \label{eq:tp}
    |t| \approx p^2 \theta^2 \approx p_T^2 = p_x^2 + p_y^2,
\end{equation}
gdzie $\vp = (p_x, p_y)$ jest pędem poprzecznym wybranej cząstki.

Związek \ref{eq:tp} jest bardzo istotny, ponieważ 
w zastosowanym układzie odniesienia składowe $p_x$ i~$p_y$ pędu poprzecznego są zmiennymi sprzężonymi do składowych $b_x$ i $b_y$ parametru zderzenia.
W ramach mechaniki kwantowej oznacza to, że amplituda rozproszenia
w reprezentacji pędowej, $T(\vp)$, jest związana z~amplitudą w~reprezentacji położeniowej%
\footnote{W artykule zastosowana została konwencja, w której to samo oznaczenie stosowane jest do matematycznie różnych, ale blisko związanych, funkcji, a ich rozróżnienie jest jednoznacznie dokonywane na podstawie argumentu.}.
Związek ten ma postać transformacji Fouriera:
\begin{equation}
    T(\vp) = \int T(\vb) e^{i \vp \cdot \vb} \text d^2 b, 
    \label{eq:fourier}
\end{equation}

W sytuacji niespolaryzowanych wiązek, amplituda $T(\vp)$ zależy jedynie od $t$, a nie zależy od $\varphi$.
Podobnie $T(\vb)$ zależy jedynie od $b$, ale nie od kierunku $\vb$.
Wygodnie wtedy stosować amplitudy będące funkcją jednej zmiennej:
\begin{equation}
    T(\vp) \to T(t), \qquad T(\vb) \to T(b).
\end{equation}
Te amplitudy związane są ze sobą przez, zbliżoną do transformacji Fouriera, transformację Hankla:
\begin{equation}
    T(t) = \int_0^\infty \text d b\, b\, T(b)\,  J_0(b\sqrt{t}).
    \label{eq:hankel}
\end{equation}

W mechanice klasycznej wartość parametru zderzenia jednoznacznie wyznacza kąt rozproszenia, czyli wartość zmiennej $t$.
Im większy parametr zderzenia, tym mniejszy kąt rozproszenia i mniejsze $|t|$.
Dokładna postać tej zależności jest konsekwencją określonej struktury wewnętrznej protonu.
W~mechanice kwantowej związek parametru zderzenia z $t$, dany wzorami (\ref{eq:fourier}) i (\ref{eq:hankel}), jest bardziej subtelny i ma charakter statystyczny.
Można natomiast powiedzieć, że małe wartości $b$ (patrząc średnio) prowadzą do dużych wartości $|t|$ (również średnio).
Nie zachodzi natomiast bezpośredni związek między kątem rozproszenia a parametrem zderzenia w poszczególnych przypadkach rozpraszania.
Informacja o strukturze przestrzennej protonu jest zawarta w amplitudzie $T(b)$.

\section{Mechanizmy oddziaływania}

\begin{figure}[hbtp]
    \includegraphics[width=\linewidth]{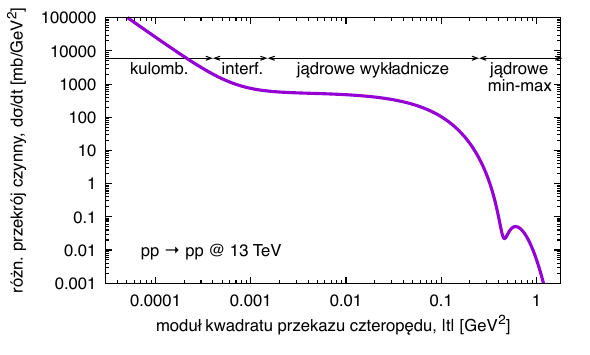}
    \caption{Przykładowy rozkład kwadratu przekazu czteropędu w procesie $pp\to pp$ przy energii zderzenia 13 TeV w~układzie środka masy.}
    \label{fig:tdist}
\end{figure}
Przykładowy rozkład zmiennej $t$ jest pokazany na rysunku \ref{fig:tdist}.
Tutaj warto wyjaśnić oznaczenie osi pionowej.
Jest to różniczkowy przekrój czynny w~$t$, $\text{d} \sigma / \text{d} t$.
Przekrój czynny jest wielkością powszechnie używaną w fizyce jądrowej oraz fizyce cząstek i~jest on niezależną od warunków eksperymentalnych miarą prawdopodobieństwa zajścia procesu fizycznego o określonym stanie początkowym i~końcowym.
Różniczkowy przekrój czynny ma się do przekroju czynnego tak jak gęstość prawdopodobieństwa do prawdopodobieństwa, tzn.
\begin{equation}
    \int_\Omega \frac{\text d \sigma}{\text d t} \text d t = \sigma_\Omega,
\end{equation}
gdzie $\Omega$ jest dowolnym podzbiorem możliwych wartości $t$, a $\sigma_\Omega$ jest przekrojem czynnym na rozpraszanie elastyczne z parametrem $t \in \Omega$.
Różniczkowy przekrój czynny związany jest z amplitudą rozproszenia
wzorem%
\footnote{Konkretna postać tego i innych wzorów użytych w artykule zależy od wybranej konwencji normalizacji amplitud.}: 
\begin{equation}
    \frac{\text d \sigma}{\text d t} = |T(t)|^2.
\end{equation}

Skomplikowany kształt rozkładu $t$ jest skutkiem dość złożonej dynamiki oddziaływań elastycznych wysokoenergetycznych protonów.
W przypadku zderzeń badanych w omawianym eksperymencie, rozpraszanie charakteryzujące się 
najmniejszymi wartościami zmiennej $t$ ($|t| \ll 7\cdot10^{-4}\ \text{GeV}^2$, co odpowiada kątom rozpraszania $\theta \ll 4$~µrad) jest przede wszystkim skutkiem odpychania ładunków elektrycznych protonów, czyli zwykłego \emph{oddziaływania kulombowskiego}. 
Wkład ten można z bardzo dobrą dokładnością obliczyć w~ramach elektrodynamiki kwantowej%
\footnote{
Należy dodać, że w literaturze rozważana jest tzw. \emph{faza kulombowska}, czyli niewielka faza zespolona, którą nabywa amplituda kulombowska jeśli cząstki oddziałują nie tylko elektromagnetycznie, ale i silnie.
Efekt ten nie jest tożsamy z omawianym dalej efektem interferencji.
Jego istnienie nie jest jednak powszechnie przyjmowane, dlatego został on pominięty we wzorze \ref{eq:TC}.
}:
\begin{equation}
\label{eq:TC}
T_C(t) = -\frac{2\sqrt{\pi}\alpha G_E^2(t)}{t},
\end{equation}
gdzie $G_E(t)$ jest elektrycznym czynnikiem kształtu (ang. \textit{electric form factor}) protonu, a $\alpha$ jest stałą struktury subtelnej.

Przy większych kątach rozproszenia ($|t| \gg 7\cdot10^{-4}\ \text{GeV}^2$) dominującym staje się mechanizm oddziaływania jądrowego silnego,
określany w~tym kontekście jako \emph{oddziaływanie jądrowe}.
Jeśli ograniczymy się dodatkowo do niezbyt dużych wartości $|t|$, $|t| \lesssim 0,1\ \text{GeV}^2$,
amplituda jądrowa daje się w dobrym przybliżeniu opisać funkcją wykładniczą:
\begin{equation}
\label{eq:TN} 
T_N(t) = A e^{-B|t|/2}
\end{equation}
co prowadzi do postaci wykładniczej różniczkowego przekroju czynnego:
\begin{equation}
\frac{\text d \sigma_N}{\text d t} = |A|^2 e^{-B|t|}.
\label{eq:dsdtN}
\end{equation}

W zakresie pośrednim, przy $|t| \sim 7\cdot10^{-4}\ \text{GeV}^2,$ amplitudy oddziaływania kulombowskiego i jądrowego mają podobną wartość bezwzględną.
Umożliwia to obserwację efektów ich interferencji. 

W zakresie $|t|$ pomiędzy 0,1 a 1 $\text{GeV}^2$ różniczkowy przekrój czynny ma charakterystyczną strukturę z lokalnym minimum i~maksimum (ang. \textit{dip and bump}).
Są one skutkiem zjawiska analogicznego do dyfrakcji światła widzialnego na małej przeszkodzie, na przykład na cienkim drucie, kiedy część padającej fali jest absorbowana, a pozostała część podlega interferencji z sobą samą. 
Z tego podobieństwa bierze się nazwa używana często dla oddziaływania elastycznego -- proces dyfrakcyjny.
W~omawianym kontekście, absorpcją są procesy nieelastyczne, czyli takie, w których stan końcowy składa się z innych cząstek niż stan początkowy, na przykład $pp \to n\pi^+\pi^0\pi^0\pi^+\pi^-p$.
Konkretny kształt struktury minimum i maximum wynika z kształtu amplitudy rozpraszania $T_N(b)$, która jest z kolei związana ze strukturą przestrzenną protonu.

Dla najwyższych wartości $|t|$, $|t| \gg 1\ \text{GeV}^2$, proces przechodzi w reżim tak zwanych oddziaływań twardych, kiedy to, paradoksalnie, oddziaływania silne stają się bardzo słabe%
\footnote{Za teoretyczne wyjaśnienie tej własności oddziaływań silnych jako rezultatu tak zwanej asymptotycznej swobody chromodynamiki kwantowej przyznana została w 2004 roku nagroda Nobla dla Dawida Grossa, H. Dawida Politzera i Franka Wilczka.}.
Oczekuje się, że w~tym reżimie proces będzie można modelować teoretycznie za pomocą  perturbacyjnej chromodynamiki kwantowej (pQCD -- ang. \textit{perturbative quantum chromodynamics}).

\section{Twierdzenie optyczne i parametr $\rho$}

\newcommand{\stot}{\ensuremath{\sigma_\text{tot}}\xspace}
Konsekwencją unitarności ewolucji układu kwantowego, jakim są oddziałujące cząstki, jest związek pomiędzy rozpraszaniem elastycznym a procesami nieelastycznymi.
W zastosowaniu do oddziaływania jądrowego ma on postać tzw. \emph{twierdzenia optycznego}:
\begin{equation}
   \stot = 4 \sqrt{\pi}\, \text Im T_N(0)
\end{equation}
gdzie
$\text Im T_N(0)$ jest urojoną częścią amplitudy rozpraszania pod kątem zero,
a \stot jest całkowitym przekrojem czynnym, czyli przekrojem czynnym na jakiekolwiek silne oddziaływanie przy tym samym stanie początkowym.
Wliczają się do niego zarówno jądrowe procesy elastyczne, jak również wszystkie możliwe jądrowe procesy nieelastyczne.

Twierdzenie optyczne pozwala wyrazić parametr $A$ ze wzoru (\ref{eq:TN}) przez całkowity przekrój czynny:
\begin{equation}
\label{eq:A}
A = \frac{(\rho + i) \stot}{4\sqrt{\pi}}.
\end{equation}
Parametr $\rho$ jest stosunkiem części rzeczywistej do urojonej jądrowej amplitudy rozpraszania pod kątem zero ($t=0$):
\begin{equation}
\label{eq:rho}
\rho = \frac{\text{Re} T_N(0)}{\text{Im} T_N(0)}.
\end{equation}

Dla $|t|$ wystarczająco dużych by móc pominąć oddziaływanie kulombowskie, ale wystarczająco małych by rozkład był z dobrym przybliżeniem wykładniczy, otrzymujemy:
\newcommand{\dsdt}{\ensuremath{\frac{\text d \sigma}{\text d t} }}
\newcommand{\dsdti}{\ensuremath{{\text d \sigma} / {\text d t} }\xspace}
\begin{equation}
   \dsdt = \frac{(1 + \rho^2)\stot^2}{16\pi}  e^{-B|t|},
\end{equation}
Ponieważ wartość $\rho$ jest rzędu 0,1, ma ona niewielki wpływ na różniczkowy przekrój czynny w tym zakresie. 
Zaniedbując ją otrzymujemy:
\begin{equation}
\stot^2 \approx 16 \pi \dsdt\Big|_{t\to0}.
\end{equation}

Rozpraszania pod kątem zero, czyli z $t=0$, nie można zmierzyć w eksperymentach na zderzających się wiązkach.
Nie można zatem bezpośrednio wyznaczyć wartości amplitudy dla $t=0$, a tym bardziej jej urojonej części. 
Można jednak wyznaczyć \stot poprzez pomiar \dsdti dla $|t|>0$ i ekstrapolację do $t=0$. 
Stanowi to jedną z głównych motywacji pomiaru opisanego w tym artykule.

Wartość całkowitego przekroju czynnego w zderzeniach proton--proton jest jedną z podstawowych własności oddziaływań silnych.
Jednak przy obecnym stanie wiedzy nie da się jej obliczyć teoretycznie z zasad pierwszych.
Wprawdzie przyjmuje się, że chromodynamika kwantowa jest poprawną teorią oddziaływań silnych, ale wciąż nie są znane metody obliczenia wartości \stot z tej teorii.
A wielkość ta pełni niezwykle istotną rolę z punktu widzenia modelowania różnorakich oddziaływań protonów przy wysokich energiach.
Modelowanie takie wykorzystuje się na przykład przy poszukiwaniu nowej fizyki, czyli zjawisk nie przewidywanych przez obecny Model Standardowy fizyki cząstek, ale też w astrofizyce cząstek, przy analizie wielkich pęków atmosferycznych wytwarzanych przez promieniowanie kosmiczne.

Pozostaje jeszcze pytanie skąd znana jest wartość parametru $\rho$.
Jak wiadomo, w mechanice kwantowej prawdopodobieństwa zajścia różnych procesów zależą jedynie od modułu amplitudy i~nie da się z nich wyznaczyć jej części rzeczywistej i urojonej.
Jeżeli jednak mamy do czynienia z interferencją dwóch amplitud, moduł ich sumy zależy od względnej zespolonej fazy pomiędzy nimi.
Fazę amplitudy można zatem wyznaczyć, mierząc interferencję z amplitudą, którą możemy obliczyć teoretycznie.
Właśnie z taką sytuacją mamy do czynienia tutaj.
Różniczkowy przekrój czynny jest dany przez koherentną sumę amplitudy kulombowskiej i jądrowej:
\begin{equation}
\label{eq:dsdt}
\dsdt = |T_C(t) + T_N(t)|^2.
\end{equation}
Interferencja jest najlepiej widoczna gdy obie amplitudy mają porównywalne wielkości, czyli 
 $|t|$ w okolicach $7\cdot 10^{-7}$~GeV$^2$. 
Pomiar $\dsdti$ w tym zakresie umożliwia wyznaczenie parametru $\rho$, a jego wartość okazuje się rzeczywiście rzędu 0,1.

Drugim sposobem na uzyskanie informacji na temat $\rho$ są relacje dyspersji.
Wynikają one z własności analityczności i unitarności amplitud rozpraszania.
Wiążą one ze sobą wartości tych amplitud przy różnych energiach zderzenia.
Między innymi wyrażają one część rzeczywistą amplitudy przez całkę%
\footnote{Całkowanie jest po energii zderzenia od progu kinematycznego do nieskończoności.}
z jej części urojonej.

\section{Eksperyment akceleratorowy}

Pomiar rozpraszania elastycznego na akceleratorze LHC nie jest łatwy.
Konieczna jest rejestracja protonów rozproszonych pod kątem nawet pojedynczych mikroradianów, czyli dziesięciotysięcznych części stopnia.
Tymczasem przy standardowych warunkach pracy akceleratora, 
wiązki w~punkcie przecięcia, czyli tam, gdzie zachodzą oddziaływania protonów, mają 
rozmycie kątowe rzędu 30~µrad.
Oznacza to, że nie da się odróżnić protonu rozproszonego elastycznie pod tak małym kątem od protonu, który nie uległ zderzeniu.
Można temu jednak zaradzić poprzez specjalne ustawienie pól w magnesach kwadrupolowych kształtujących wiązki.
Jest to tak zwana specjalna optyka akceleratora.
Używa się tutaj nazwy \emph{optyka}, ponieważ magnesy kwadrupolowe w~akceleratorze działają podobnie jak soczewki -- skupiają i rozpraszają wiązkę.

W standardowej optyce akceleratora dąży się do maksymalnego skupienia wiązek w~punkcie zderzenia, aby zmaksymalizować szansę na zaobserwowanie rzadkich procesów.
Skutkuje to jednak dużym rozmyciem kątowym wiązek.
Jeśli wiązki skupi się mniej, rozmycie kątowe będzie mniejsze.
Zwiększy się za to przekrój poprzeczny wiązek.
W~LHC rozmiar ten jest typowo rzędu 10~µm.
Przy specjalnej optyce wynosi niemalże milimetr, a rozmycie kątowe maleje do 0,2~µrad. 

\begin{figure}[t]
    \centering
    \resizebox{0.9\linewidth}{!}{%
    \begin{tikzpicture}[scale=0.25,thick,yscale=-1]
\tikzstyle{every node}=[font=\LARGE]

\fill [black!10] (1,4) rectangle (18,12);
\fill [black!10] (5.4,9) rectangle (13.6,15);

\draw[blue!80!black,fill=blue!20]  (6.5,9) rectangle (12.5,15);
\draw[green!50!black,fill=green!60!black!30]  (7,9.3) rectangle (12,12);
\draw [] (5,12) -- (1,12);
\draw [] (14,12) -- (18,12);
\begin{scope}[rotate around={90:(14,12)}]
\foreach \x in {0,...,8}{
  \draw  (14+\x*0.33222591362126247,12) -- ++(0.16611295681063123,1) -- ++ (0.16611295681063123, -1);
}
\end{scope}
\begin{scope}[rotate around={-90:(5,15)}]
\foreach \x in {0,...,8}{
  \draw  (5+\x*0.33222591362126247,15) -- ++(0.16611295681063123,1) -- ++ (0.16611295681063123, -1);
}
\end{scope}
\draw [] (5,15) -- (14,15);
\draw [] (1,4) -- node [below] {\tiny rura akceleratorowa}(18,4) ;
\draw [red!80!black, line width = 5pt] (1,8) -- (18,8) node [below, pos=0.9] {\tiny wiązka};
\draw[gray,<->] (15,13) -- node[right] {\tiny ruch} +(0,3);

\node[blue!80!black] at (9.5,14.3) {\tiny ,,garnek''};
\node[green!50!black] at (9.5,11.3) {\tiny detektor};
\end{tikzpicture}%
    }%
    \caption{Schemat rzymskiego garnka.}
    \label{fig:romanpot}
\end{figure}

\begin{figure*}[t!]
    \centering
    \resizebox{0.9\linewidth}{!}{%
    \tiny\usetikzlibrary {arrows.meta}
\usetikzlibrary {calc}
\begin{tikzpicture}[scale=0.25,thick]
\newcommand{\OZ}{14} 
\newcommand{\IZ}{12} 
\newcommand{\ST}{4} 
\newcommand{\SB}{1} 
\newcommand{\DIST}{-5} 
\newcommand{\MI}{4} 
\newcommand{\MO}{9} 
\newcommand{\MS}{4} 
\newcommand{\CM}{green!70!black} 
\newcommand{\CP}{blue!70!black} 
\newcommand{\BZ}{16} 
\newcommand{\LA}{18} 
\draw [red!70!black] (-\BZ, 0) -- node[pos=1.0,above] {wiązka} (\BZ,0);
\draw [ultra thick] (-\OZ, -\ST) -- (-\OZ, -\SB);
\draw [ultra thick] (-\OZ,  \ST) -- (-\OZ,  \SB);
\draw [ultra thick] ( \OZ, -\ST) -- ( \OZ, -\SB);
\draw [ultra thick] ( \OZ,  \ST) -- ( \OZ,  \SB);
\draw [ultra thick] (-\IZ, -\ST) -- (-\IZ, -\SB);
\draw [ultra thick] (-\IZ,  \ST) -- (-\IZ,  \SB);
\draw [ultra thick] ( \IZ, -\SB) -- ( \IZ, -\ST);
\draw [ultra thick] ( \IZ,  \ST) -- ( \IZ,  \SB);

\draw [gray,thin,Stealth-Stealth] ( \IZ,  \DIST) -- node[below] {237 m} ( 0,  \DIST);
\draw [gray,thin,Stealth-Stealth] (-\IZ,  \DIST) -- node[below] {237 m} ( 0,  \DIST);
\draw [gray,thin,Stealth-Stealth] (-\IZ,  \DIST) -- node[below] {8 m} (-\OZ,  \DIST);
\draw [gray,thin,Stealth-Stealth] ( \IZ,  \DIST) -- node[below] {8 m} ( \OZ,  \DIST);

\draw [gray,thin,Stealth-Stealth] ( -\LA,  -\SB) -- node[left] {2 mm} ( -\LA,  \SB);
\draw [gray,thin,Stealth-Stealth] ( -\LA,  \SB) -- node[left] {20 mm} ( -\LA,  \ST);
\draw [gray,thin,Stealth-Stealth] ( -\LA, -\SB) -- node[left] {20 mm} ( -\LA, -\ST);

\fill [\CM, opacity=0.2] (\MI, \MS) -- (\MO, \MS) -- (\MO, -\MS) -- (\MI, -\MS) -- cycle;
\fill [\CM, opacity=0.2] (-\MI, \MS) -- (-\MO, \MS) -- (-\MO, -\MS) -- (-\MI, -\MS) -- cycle;
\draw [\CP,->] (0,0) [rounded corners] -- (\MI,  0.3) -- ($(\MI,0.3)!0.6!(\MO, 1.0) $) -- ( \MO, 1.3) -- ( \BZ, 4);
\draw [\CP,->] (0,0) [rounded corners] -- (-\MI,-0.3) --  ($(-\MI,-0.3)!0.6!(-\MO, -1.0) $) -- (-\MO,-1.3) -- (-\BZ,-4);

\fill [\CM] (\MI, \MS) -- node[above] {magnesy} (\MO, \MS);
\fill [\CM] (-\MI, \MS) -- node[above] {magnesy} (-\MO, \MS);
\draw [] (0,0) circle (0.1); 
   
\path [] (\OZ, \ST) -- node[above] {detektory} (\IZ, \ST);
\path [] (-\OZ, \ST) -- node[above] {detektory} (-\IZ, \ST);

\draw[->] (16, -4) -- +(2, 0) node[right] {$z$};
\draw[->] (16, -4) -- +(0, 2) node[above] {$y$};
\end{tikzpicture}%
    }%
    \caption{Schemat układu eksperymentalnego detektorów ATLAS--ALFA.
    Kolorem czarnym zaznaczono stacje detektorowe, kolorem zielonym -- magnesy kwadrupolowe, kolorem czerwonym -- wiązkę protonową, kolorem niebieskim -- protony rozproszone elastycznie.
    }
    \label{fig:exp}
\end{figure*}

\begin{figure*}[htbp]
\centering
\includegraphics[width=\linewidth]{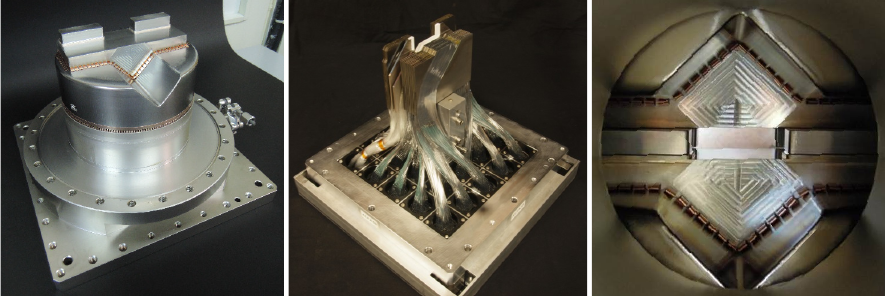}
    \caption{Zdjęcia elementów układu ALFA. Od lewej: rzymski garnek, detektor scyntylacyjny wkładany do garnka, widok dwóch detektorów -- górnego i dolnego -- z perspektywy wiązki protonowej. Źródło: \cite{ALFA} oraz materiały własne Współpracy ATLAS-ALFA.}
    \label{fig:rpphotos}
\end{figure*}

Sama możliwość odróżnienia protonu rozproszonego od protonów wiązki nie wystarcza.
Trzeba jeszcze taki rozproszony proton zarejestrować.
Tu pojawia się kolejna trudność.
Ponieważ kąt rozproszenia jest bardzo mały, proton pozostaje wewnątrz rury próżniowej akceleratora.
Co więcej, ponieważ wiązki są teraz szerokie, przez długi czas znajduje się on wewnątrz samej wiązki.

Aby móc zarejestrować takie protony, korzysta się z~techniki tak zwanych rzymskich garnków%
\footnote{Nazwa pochodzi od grupy fizyków z Rzymu, która jako pierwsza zastosowała tę technikę w latach 70-tych w akceleratorze ISR w CERN, oraz od kształtu głównej części urządzenia przypominającej stalowy garnek kuchenny.}.
Detektory umieszczane są wewnątrz rury akceleratora w stalowym naczyniu.
Precyzyjny silnik krokowy umożliwia przysuwanie detektora w pobliże wiązki na czas pomiaru.
Wcześniej, to znaczy gdy wiązka jest wstrzykiwana do akceleratora i rozpędzana, 
detektory są odsunięte na bezpieczną odległość.

Stalowy ,,garnek'' stanowi ochronę ultrawysokiej próżni akceleratora odgradzając ją od wtórnej próżni otaczającej detektory.
Wtórna próżnia w niewielkiej objętości rzymskiego garnka nie musi być aż tak wysokiej jakości jak główna próżnia i znacznie łatwiej ją uzyskać.
Garnek umożliwia również fizyczny dostęp do detektorów bez naruszania głównej próżni.
Schemat rzymskiego garnka pokazany został na rysunku \ref{fig:romanpot}.

Detektory ATLAS-ALFA \cite{ALFA}, czyli układ eksperymentalny omawianego pomiaru,
zaprezentowany został na rysunku~\ref{fig:exp}.
Składa się on z czterech stacji.
Każda stacja składa się z~dwóch rzymskich garnków: jednego dosuwanego do wiązki od~dołu i drugiego dosuwanego od góry. 
Aby móc precyzyjnie mierzyć bardzo małe kąty rozproszenia protonów, detektory umieszczone zostały w odległości ponad 200 metrów od punktu zderzenia i~dosunięte na zaledwie jeden milimetr od wiązki.

Detektor ma kilka części. 
Najważniejszą z nich jest detektor główny, którego zadaniem jest pomiar położenia przelatujących protonów.
Składa się on z~20 warstw włókien scyntylacyjnych.
Włókna mają przekrój kwadratowy o boku długości 0,5~mm.
Cząstka naładowana przelatująca przez takie włókno wytwarza w nim impulsy światła (scyntylacje),
które propagując się wzdłuż włókna docierają do wielokanałowego fotopowielacza.
Sygnał z każdego kanału jest odczytywany i digitalizowany przez elektronikę odczytu.
Wiedząc które włókna zostały ,,zapalone'', dostajemy informację o położeniu cząstki.
Na rysunku \ref{fig:rpphotos} pokazane zostały zdjęcia elementów użytego układu doświadczalnego.

Wspomniana wcześniej optyka akceleratora jest kluczowa nie tylko dla samej możliwości pomiaru, ale również przy wyznaczaniu kąta rozproszenia ze zmierzonego położenia protonu.
Gdyby pomiędzy punktem zderzenia a detektorem nie było żadnych pól, współrzędne położenia poprzecznego protonu w detektorze $(x,y)$ byłyby równe:
\begin{align} 
\label{eq:transportnofield}
x &=  x_0 + L \tg\theta_x \approx x_0 + L\cdot \theta_x, \nonumber\\
y &=  y_0 + L \tg\theta_y \approx y_0 + L\cdot \theta_y,
\end{align}
gdzie 
$(x_0, y_0)$ to współrzędne poprzeczne w miejsca oddziaływania,
$L$ to odległość wzdłuż osi $z$ pomiędzy punktem oddziaływania i detektorem,
a kąty $\theta_x$ i $\theta_y$ dane są wzorami%
\footnote{
Ściśle rzecz biorąc, z uwagi na rozmycie kątowe zderzających się wiązek, kąty $\theta$, $\varphi$ w tych wzorach są kątami nachylenia trajektorii w punkcie zderzenia, a nie bezpośrednio kątami rozproszenia.
Dla uproszczenia rozróżnienie to zostało tu pominięte.
}:
\begin{align} 
\theta_x = \theta \cos \varphi \nonumber \\
\theta_y = \theta \sin \varphi.
\end{align}

Gdy trajektoria protonu przechodzi przez pole magnetyczne, równania (\ref{eq:transportnofield}) nie są już prawdziwe.
Rozwiązując równania ruchu w polu magnetycznym okazuje się jednak, że 
trajektorie protonów rozproszonych elastycznie spełniają podobne, ale ogólniejsze równania:
\newcommand{\Leffx}{\ensuremath{L_{\text{eff}, x}}\xspace}
\newcommand{\Leffy}{\ensuremath{L_{\text{eff}, y}}\xspace}
\begin{align} 
\label{eq:transportfield}
x &=  \alpha_x x_0 + \Leffx\cdot \theta_x, \nonumber\\ 
y &=  \alpha_y x_0 + \Leffy\cdot \theta_y.
\end{align}
Parametry \Leffx i \Leffy są często określane mianem odległości efektywnej i razem z parametrami $\alpha_x$ i $\alpha_y$ zależą od zastosowanej optyki.
W omawianym pomiarze odległość geometryczna $L$ dalszych stacji wynosi 245~m,
natomiast odległości efektywne mają w~przybliżeniu wartości $\Leffx = 33$~m i~$\Leffy = 320$~m.
Warto zwrócić uwagę na znaczącą różnicę pomiędzy $\Leffx$ i $\Leffy$.
Wynika ona z~własności magnesów kwadrupolowych, które skupiając wiązkę w kierunku $x$ rozpraszają ją w kierunku $y$, i na odwrót.
\section{Analiza danych}

Analiza danych jest oparta na własnościach rozpraszania elastycznego.
Jak wspomniano wcześniej, kinematyka pojedynczego przypadku takiego oddziaływania jest w pełni scharakteryzowana przez dwa parametry: $\theta$ i $\varphi$ (alternatywnie: $\theta_x$ i~$\theta_y$).
Tymczasem zastosowany układ doświadczalny pozwala na niezależny pomiar ośmiu wielkości charakteryzujących
każdy przypadek, tzn. współrzędnych $x$ i $y$ w~każdej z~czterech stacji detektorowych (por. rys. \ref{fig:exp}).
Mierzone współrzędne muszą zatem być silnie skorelowane.

\begin{figure}[htbp]
    \centering
    \includegraphics[width=\linewidth]{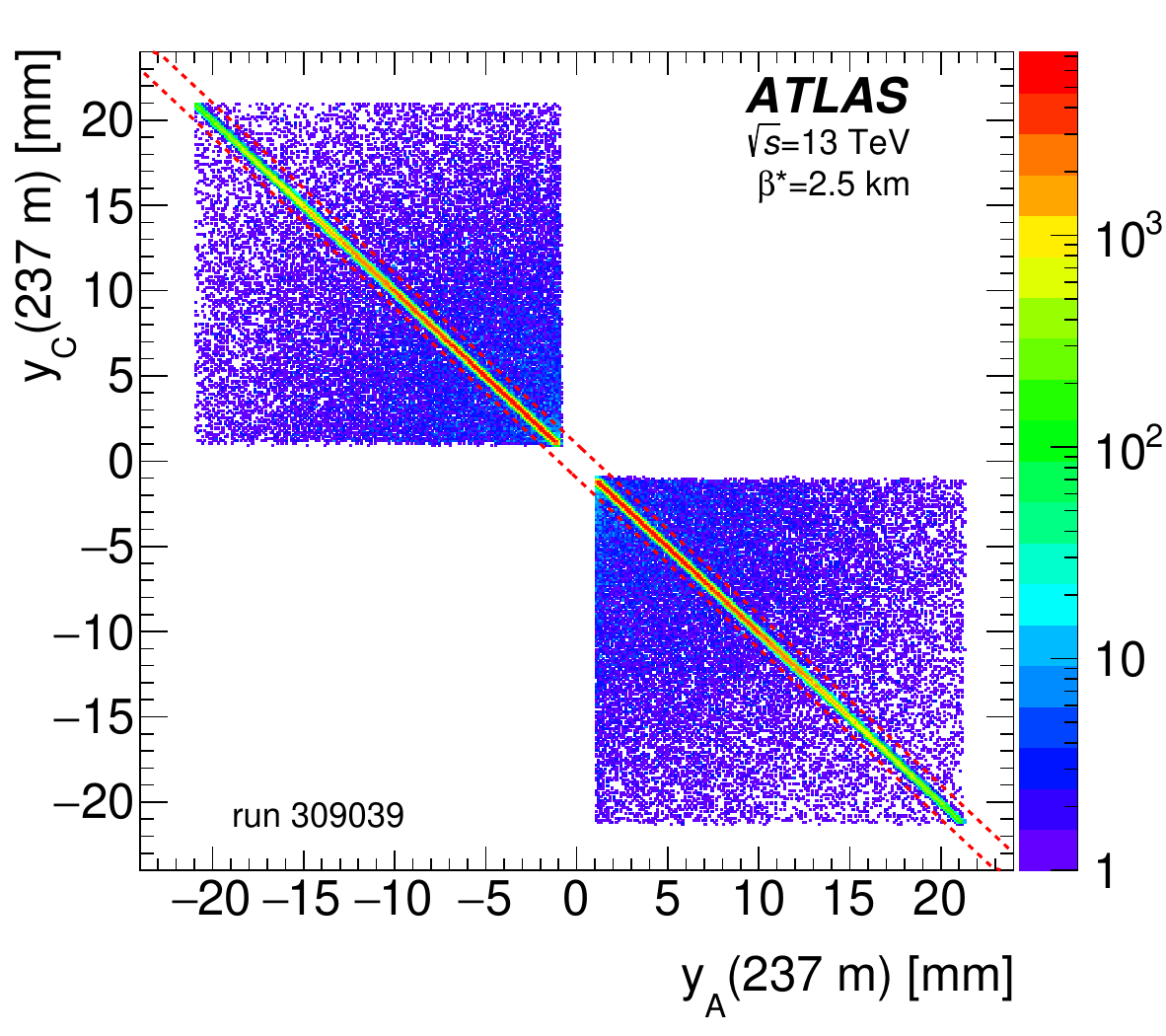}
    \caption{Korelacja pomiędzy współrzędną pionową położenia protonów zarejestrowanych po prawej ($y_C$) i lewej ($y_A$) stronie punktu oddziaływania w detektorach oddalonych o 237~m.
    Czerwone przerywane linie wskazują zastosowaną selekcję.
    Skala kolorów wskazuje liczbę zarejestrowanych przypadków.
    Źródło: \cite{STDM-2018-08}.
    }
    \label{fig:ycorrelation}
\end{figure}

Najprostszą korelacją jest ta występująca pomiędzy położeniami obu mierzonych protonów.
Ponieważ układ odniesienia związany z systemem detektorów jest w dobrym przybliżeniu układem środka masy zderzenia, dwa rozproszone protony mają przeciwne pędy, a współrzędne położenia mierzone po lewej i prawej stronie będą miały zbliżone wartości, ale przeciwne znaki.
Taka bardzo silna antykorelacja składowych $y$ położenia obu protonów pokazana została na rysunku \ref{fig:ycorrelation}.
Jest to korelacje, a nie ścisła zależność, ponieważ dochodzą niepewności pomiarowe, które mogą być niezależne dla każdej mierzonej wielkości.

Korelacje takie wykorzystuje się aby wybierać przypadki rozpraszania elastycznego, a odrzucać tło, czyli przypadki, które na pierwszy rzut oka mogą sprawiać wrażenie, że pochodzą z rozpraszania elastycznego, ale w istocie takimi nie są.
Źródłem tła mogą być procesy nieelastyczne oraz tak zwane \textit{halo wiązki}, czyli protony, które znajdują się daleko od jej środka.

Antykorelacje położeń protonów mierzonych po lewej i prawej stronie mogą być wykorzystane również przy wyznaczaniu kąta rozproszenia. 
Oba protony mają przeciwnie skierowane pędy, ale to samo początkowe położenie.
Biorąc różnicę mierzonych składowych położenia po obu stronach, wkład od położenia początkowego  eliminuje się.
Przykładowo dla współrzędnej $x$:
\begin{equation}
    x_L - x_R  
    = 2 \Leffx \theta_x
\end{equation}
Z pomiaru położenia obu rozproszonych protonów można więc odzyskać kąt rozproszenia.

\begin{figure}[t]
    \centering
    \includegraphics[width=\linewidth]{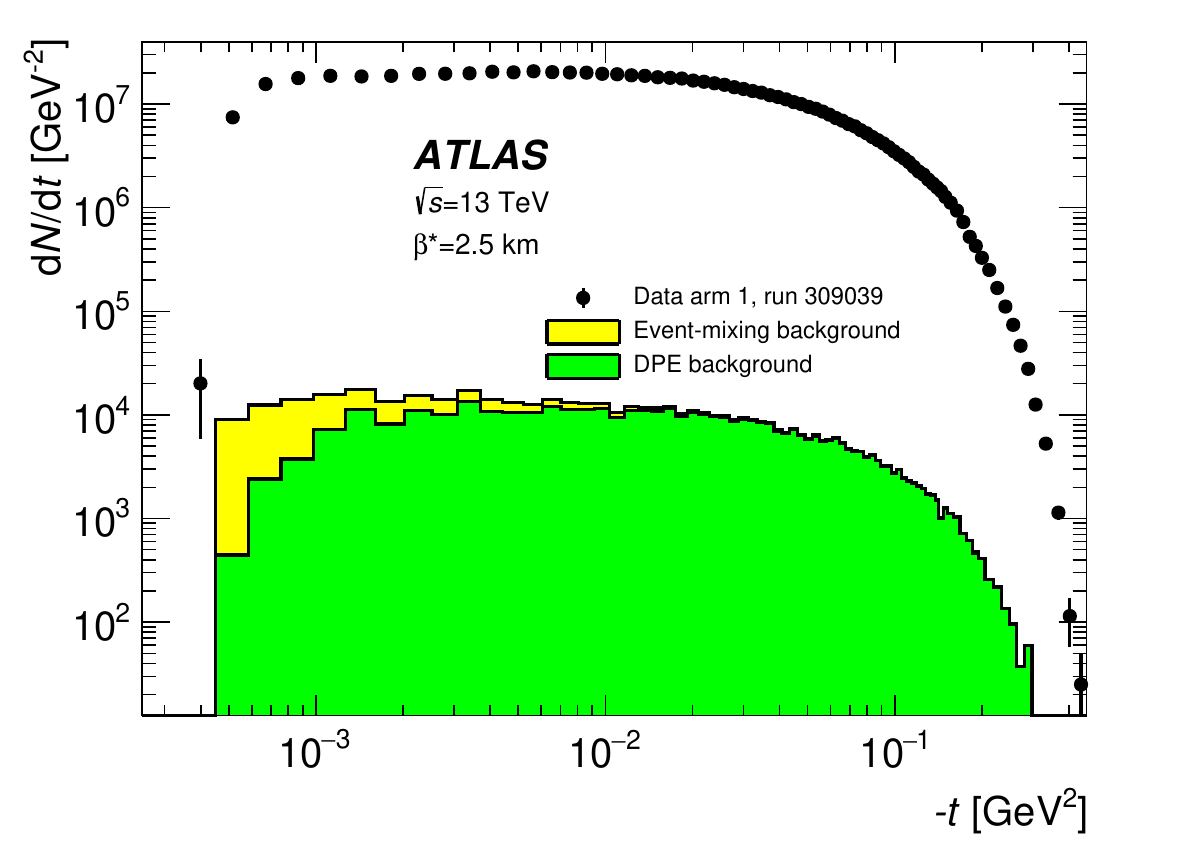}
    \caption{Rozkład $t$ uzyskany bezpośrednio z~pomiaru.
    Źródło: \cite{STDM-2018-08}.
    }
    \label{fig:rawdist}
    \includegraphics[width=\linewidth]{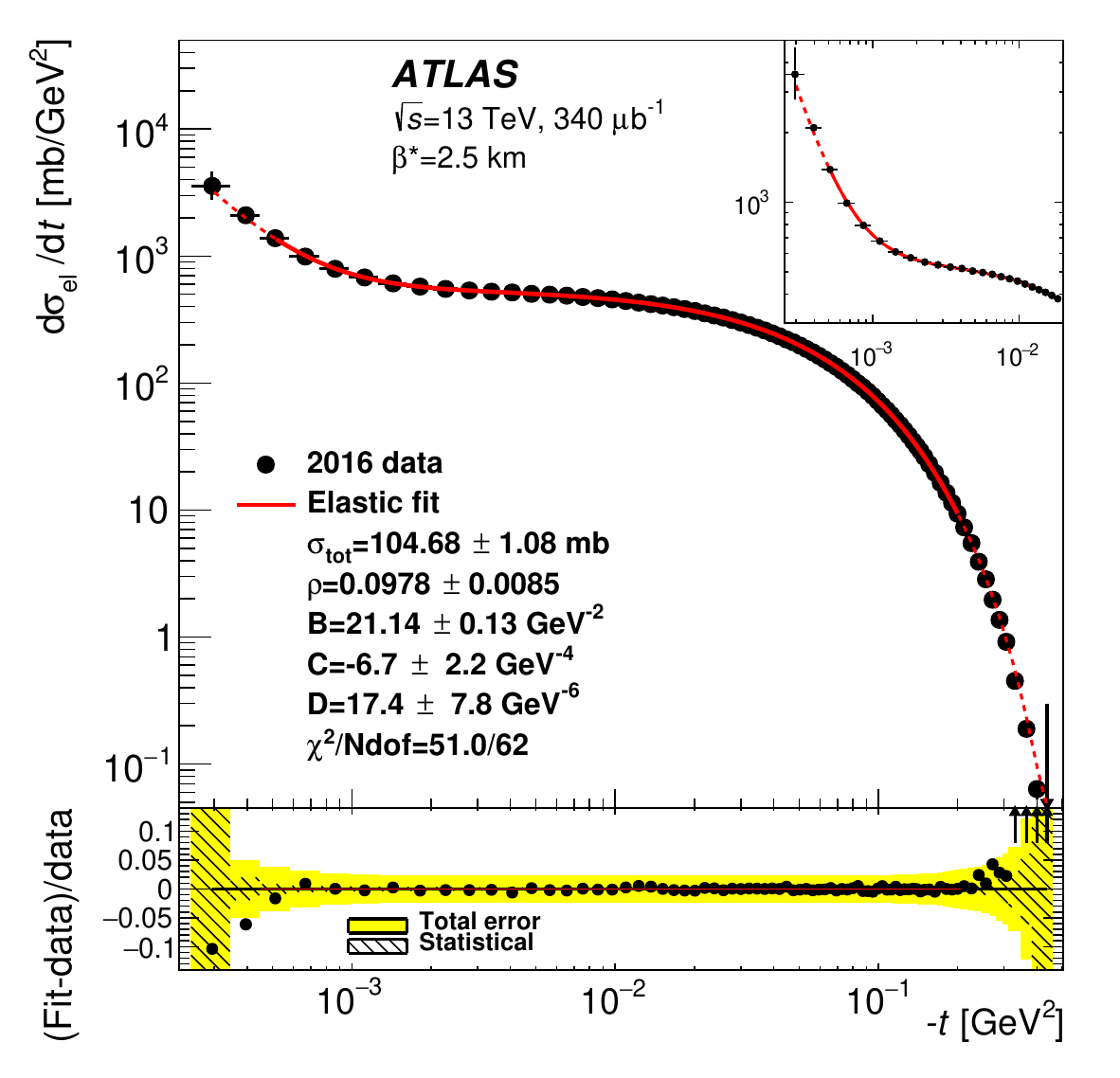}
    \caption{
    Rozkład $t$ po wszystkich poprawkach.
    Źródło: \cite{STDM-2018-08}.
    }
    \label{fig:corrections}
\end{figure}

Mając wyznaczony kąt rozproszenia i znając energię wiązki, możemy ze wzoru (\ref{eq:t}) wyliczyć wartość zmiennej $t$.
Zestawiając wszystkie zarejestrowane i zakwalifikowane jako sygnał przypadki
można utworzyć rozkład tej zmiennej.
Rysunek \ref{fig:rawdist} pokazuje taki właśnie rozkład uzyskany przy wykorzystaniu detektorów ATLAS-ALFA.
Otrzymany wynik znacząco różni się od rozkładu prawdziwego, czyli takiego, jaki uzyskano by dysponując idealnym układem pomiarowym oraz nieskończoną liczbą przypadków.
W omawianym pomiarze głównym źródłem błędów są niedoskonałości układu, a~więc efekty systematyczne. 
Są różne źródła tych niedoskonałości:
\begin{enumerate}
    \item Jeśli proton znajduje się zbyt blisko wiązki, albo jeśli uderzy w rurę lub inny element akceleratora zanim doleci do detektora, nie będzie mógł być zarejestrowany. Ten efekt nazywany jest akceptancją geometryczną.
    \item Nawet jeżeli przypadek był w akceptancji detektora, mógł nie zostać prawidłowo zarejestrowany z powodu zawsze obecnej, choć niewielkiej, niewydajności.
    \item Kąt rozproszenia mierzony jest ze skończoną zdolnością rozdzielczą. Wynika to zarówno ze skończonej przestrzennej zdolności rozdzielczej samych detektorów, 
    jak i z niezerowej rozbieżności kątowej wiązek.
    Skończona zdolność rozdzielcza pojedynczego pomiaru powoduje z kolei systematyczne wypłaszczenie mierzonego rozkładu%
    \footnote{%
    W sytuacji skończonej zdolności rozdzielczej, na pomiar rozkładu dla ustalonej wartości $t_0$ ma wpływ otoczenie $t_0$.
    Matematycznie należy tu myśleć o konwolucji rozkładu prawdziwego ze zdolnością rozdzielczą,
    skutkujące jego rozmyciem.
    }.
    Dla stromych rozkładów, a taki właśnie jest mierzony rozkład $t$, efekt ten może stać się dość istotny.
    \item 
    Zarówno położenie detektorów względem wiązki jak i pola magnetyczne, z których wyznacza się wielkość parametrów $\Leffx$ i $\Leffy$, nie są znane absolutnie dokładnie.
    \item Zawsze pozostaje jakaś część tła, której nie udało się odrzucić.
    \item Aby z liczby zebranych przypadków rozpraszania otrzymać przekrój czynny, konieczna jest znajomość świetlności akceleratora scałkowanej po okresie prowadzenia pomiarów.
    Scałkowana świetlność to miara ilości zebranych danych i~jest zależna od rozmiarów i prądów zderzających się wiązek oraz czasu zbierania danych.
    Jej wartość jest wynikiem osobnego pomiaru i jako taka jest znana ze skończoną dokładnością.
\end{enumerate}

Celem pomiaru jest uzyskanie informacji o rozkładzie prawdziwym.
Wpływ błędów systematycznych można w dużej części precyzyjnie określić z użyciem symulacji komputerowych.
Znając ten wpływ, na zmierzony rozkład nakłada się poprawki mające te błędy zniwelować.
Jeżeli precyzyjne określenie błędu nie jest możliwe, szacuje się jego prawdopodobną wielkość i uwzględnia w niepewności wyniku końcowego.

Rysunek \ref{fig:corrections} pokazuje rozkład po wszystkich poprawkach, czyli różniczkowy przekrój czynny w $t$.
Porównanie z rozkładem niepoprawionym (rysunek \ref{fig:rawdist}) pokazuje, że uwzględnienie wszystkich poprawek jest absolutnie kluczowe.

\section{Wyniki i ich interpretacja}

Aby ze zmierzonego rozkładu wydobyć interesujące własności oddziaływań silnych wykonuje się dopasowanie modelu do danych.
Zastosowany model jest dany wzorem (\ref{eq:dsdt}).
Występująca w nim amplituda $T_C$ jest dana wyrażeniem (\ref{eq:TC}) i nie ma żadnych wolnych parametrów, a amplituda
$T_N$ jest dana wyrażeniem wynikającym ze wzorów (\ref{eq:TN}) i (\ref{eq:A}):
\begin{equation}
   T_N(t) = \frac{\stot(\rho+i)}{4 \sqrt{\pi}} e^{-B|t|/2}.
\end{equation}
Wolnymi, podlegającymi dopasowaniu, parametrami są tu: \stot, $B$ i $\rho$.
Warto zwrócić uwagę na zależność od $\rho$.
Różniczkowy przekrój czynny dany wzorem (\ref{eq:dsdt}) ma trzy człony: człon kulombowski (dany przez $|T_C|^2$), człon jądrowy (dany przez $|T_N|^2$) oraz człon interferencyjny pomiędzy $T_C$ i~$T_N$.
Człon kulombowski jest niezależny od parametru $\rho$.
Człon jądrowy jest proporcjonalny do $1 + \rho^2$, a ponieważ $\rho^2 \ll 1$, wpływ $\rho$ jest tu niewielki.
Natomiast człon interferencyjny jest do $\rho$ wprost proporcjonalny.
Obecność tego właśnie członu odpowiada za czułość różniczkowego przekroju czynnego na wartość parametru $\rho$.

Dopasowanie wykonywane jest metodą najmniejszych kwadratów z uwzględnieniem korelacji pomiędzy niepewnościami poszczególnych punktów.
Na przykład cały rozkład jest wprost proporcjonalny do świetlności, więc niepewność związana z wyznaczeniem tej wielkości wpływa na wszystkie punkty rozkładu w jednakowy sposób.
Korelacja między niepewnościami różnych punktów jest stuprocentowa.
Nieuwzględnienie tej korelacji miałoby znaczący wpływ na uzyskany wynik dopasowania, w szczególności na niepewność parametru $\stot$.
Podobne, choć już nie tak silne, korelacje występują dla wszystkich źródeł niepewności systematycznej.

Najważniejszym wynikiem są uzyskane z dopasowania wartości \stot i $\rho$ dla oddziaływań $pp$ przy energii 13 TeV w układzie środka masy:
\begin{align*}
\stot &= 104,7 \pm 1.1 \ \text{mb}, \\
\rho &= 0,098 \pm 0,011.
\end{align*}
Głównym źródłem końcowego błędu pomiarowego $\stot$ są niepewności systematyczne związane z~wyznaczaniem całkowitej świetlności oraz położenia detektorów względem wiązki.
Wyznaczona wartość parametru $\rho$ zależy również od konkretnego wyboru modelu amplitudy jądrowej $T_N(t)$.

\begin{figure}[t!]
    \centering
    \includegraphics[width=\linewidth]{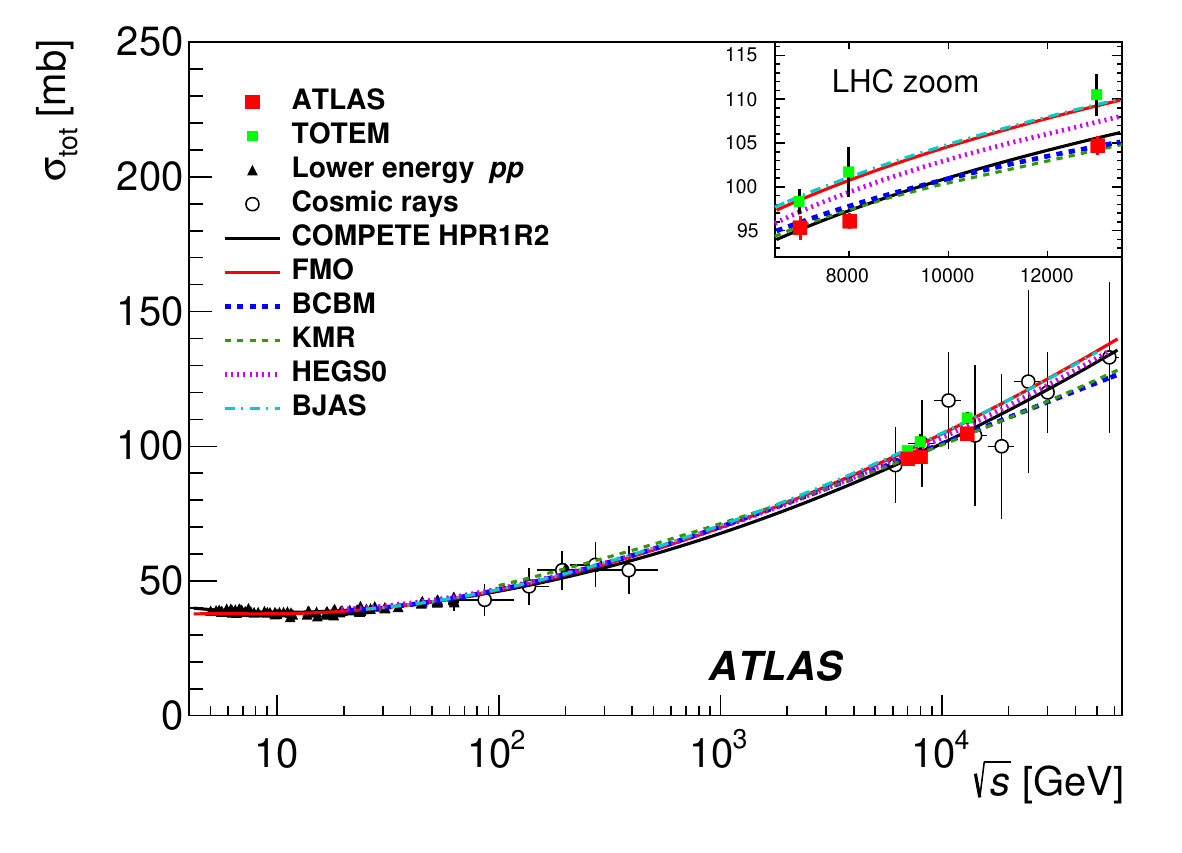}
    \includegraphics[width=\linewidth]{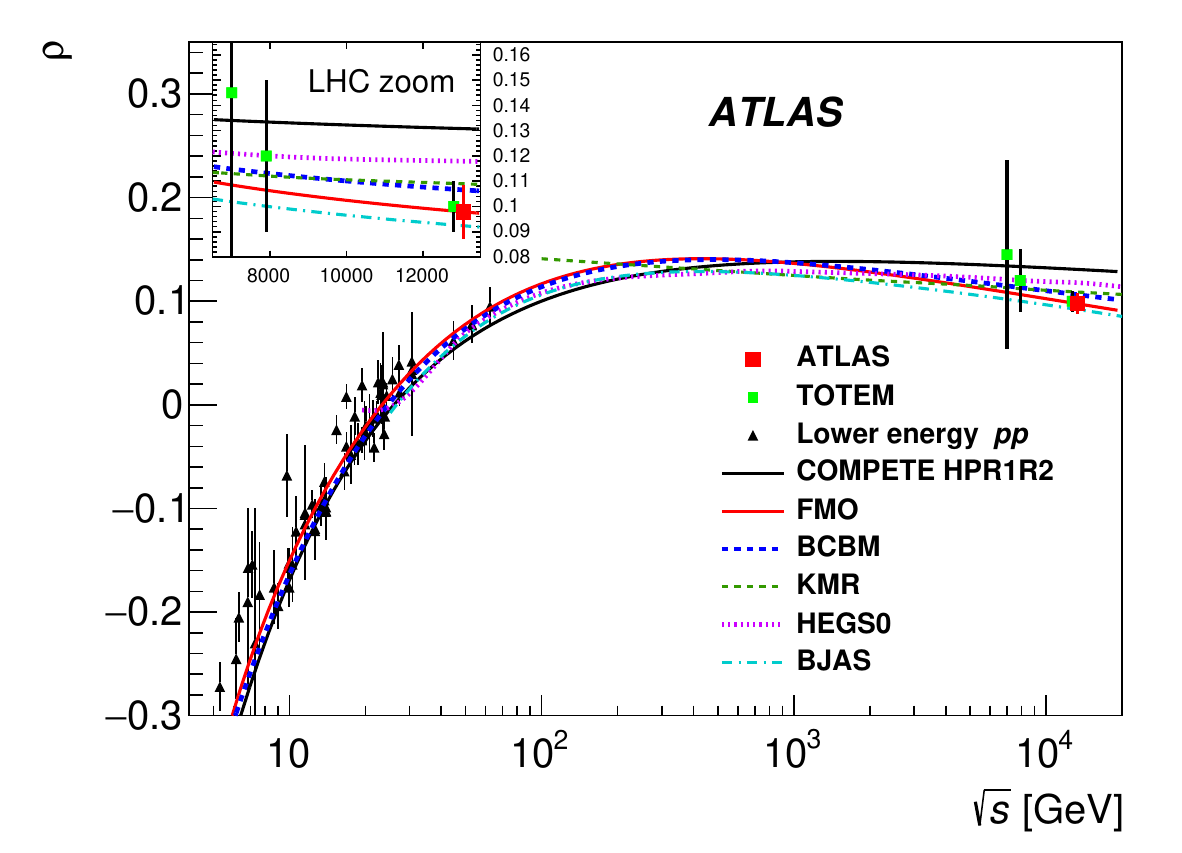}
    \caption{Wyniki pomiarów całkowitego przekroju czynnego $\stot$ (górny wykres) oraz parametru $\rho$ (dolny wykres) w~funkcji energii zderzenia w układzie środka masy, $\sqrt{s}$. wraz z przewidywaniami różnych modeli teoretycznych.
    Źródło: \cite{STDM-2018-08}.
    }
    \label{fig:results}
\end{figure}

Rysunek \ref{fig:results} pokazuje zestawienie powyższych wyników z innymi pomiarami oraz z modelami teoretycznymi.
Pierwszą interesującą obserwacją jest, że całkowity przekrój czynny rośnie wraz z energią zderzenia.
Jest to fakt znany i stosunkowo dobrze opisywany przez różne modele teoretyczne.
Można o tym efekcie myśleć jak o puchnięciu protonu.
Jest to spowodowane tym, że przy wyższej energii większe jest skrócenie Lorentza protonów w kierunku ich ruchu, przez co zderzenie trwa krócej.
Im krótszy czas oddziaływania, tym bardziej jest ono czułe na krótkotrwałe fluktuacje kwantowe -- powstawanie wirtualnych kwarków, antykwarków i gluonów wokół protonu.

Drugim ważnym wnioskiem, jaki można wyciągnąć porównując pokazane na rysunku \ref{fig:results} dane i~przewidywania teoretyczne jest to, 
że model COMPETE \cite{COMPETE} nie może jednocześnie opisać zmierzonych wartości całkowitego przekroju czynnego \stot i parametru $\rho$.
Model ten jest oparty o dwa kluczowe założenia:
\begin{enumerate}
    \item Dla wysokich energii zderzenia ($\sqrt{s} > 1$~TeV), \stot rośnie ze wzrostem energii jak jej logarytm do kwadratu\footnote{
    Wysycając tak zwane ograniczenie Froissarta.
    }.
    \item Przy energii dążącej do nieskończoności oddziaływania silne protonów z protonami oraz protonów z antyprotonami stają się takie same, to znaczy całkowite przekroje czynne w oddziaływaniach $pp$ i $p\bar p$ stają się równe asymptotycznie.
\end{enumerate}
Niezgodność wyniku doświadczalnego z modelami opartymi na tych założeniach oznacza, że (a) wzrost $\stot$ z energią zacznie zwalniać, lub że (b) oddziaływanie $pp$ różni się asymptotycznie od oddziaływań $p\bar p$.
W tym drugim przypadku słowem kluczowym jest \emph{odderon}.

\section{Odderon}

W ramach perturbacyjnej kwantowej teorii pola,
reakcje zachodzące pomiędzy cząstkami można interpretować jako skutek wymieniania między nimi innych cząstek.
Na przykład rozpraszanie elektromagnetyczne pomiędzy elektronem a mionem zachodzi dzięki wymianie fotonu.
Anihilacja pary $e^+ e^-$ w parę fotonów to z kolei wymiana elektronu.
Procesy te przedstawione zostały na rysunku \ref{fig:Feynman} w~postaci diagramów Feynmana.

\begin{figure}[h]
    \centering
\vspace{2ex}
    \resizebox{\linewidth}{!}{%
\begin{fmffile}{diagrams1}
 \begin{fmfgraph*}(80,70)
   \fmfleft{i1,i2}
   \fmfright{o1,o2}
   \fmf{fermion}{i1,v1,o1}
   \fmf{fermion}{i2,v2,o2}
   \fmf{photon,label=$\gamma$}{v1,v2}
   \fmflabel{$e^-$}{i2}
   \fmflabel{$\mu^-$}{i1}
   \fmflabel{$e^-$}{o2}
   \fmflabel{$\mu^-$}{o1}
 \end{fmfgraph*}
 \hspace{7mm}
 \begin{fmfgraph*}(80,70)
   \fmfleft{i1,i2}
   \fmfright{o1,o2}
   \fmf{fermion}{i1,v1,v2,i2}
   \fmf{photon}{o1,v1}
   \fmf{photon}{o2,v2}
   \fmflabel{$e^+$}{i2}
   \fmflabel{$e^-$}{i1}
   \fmflabel{$\gamma$}{o2}
   \fmflabel{$\gamma$}{o1}
 \end{fmfgraph*}
\end{fmffile}
}

\vspace{7mm}

\caption{Przykładowe diagramy Feynmana: 
$e\mu\to e\mu$ (z lewej), 
$e^+e^- \to \gamma\gamma$ (z prawej).
Linie po lewej stronie każdego z diagramów reprezentują cząstki w stanie początkowym, linie po prawej stronie -- w stanie końcowym.
}
\label{fig:Feynman}

\end{figure}

Wydawałoby się naturalne, by opis opis oddziaływania jako wymiany zastosować również do zderzeń hadronów, czyli cząstek oddziałujących silnie: protonów, neutronów, pionów, kaonów, \dots\ 
Na~przykład reakcja $\pi^+ n \to \pi^0 p$ mogłaby zachodzić poprzez wymianę pionu $\pi^+$. 
Jednak hadronów mogących pośredniczyć w takiej reakcji jest wiele, w szczególności mogą to być cząstki o wysokich spinach.
Model teoretyczny oparty o takie wymiany nie jest jednak poprawny, ponieważ występuje w nim naruszenie unitarności.
Okazuje się jednak, że problem znika, jeśli zamiast wymiany pojedynczej cząstki w obliczeniach uwzględni się wymianę całej rodziny nieskończenie wielu cząstek o podobnych własnościach -- tych samych liczbach kwantowych, ale różnych spinach i masach.
Taką właśnie rodzinę cząstek nazywa się trajektorią.

Przy wysokich energiach zderzających się cząstek, takich jak osiągane w LHC, wiodący wkład do rozpraszania protonów wnosi trajektoria \emph{pomeronu}.
W~przeciwieństwie do innych trajektorii, opartych o znane z eksperymentów hadrony, istnienie pomeronu zostało zapostulowane teoretycznie, aby wyjaśnić wspomniany wcześniej wzrost całkowitego przekroju czynnego z energią zderzenia.
Współcześnie przyjmuje się się, że trajektorię pomeronu tworzą cząstki zwane \emph{glueballami} lub \emph{gluoniami} -- hadronami składającymi się jedynie z gluonów (nie licząc powstających na chwilę w wyniku fluktuacji  kwantowych par kwark--antykwark).
W chwili obecnej nie mamy jasnej odpowiedzi dotyczącej istnienia i własności glueballi, więc tym bardziej ich związek z pomeronem nie jest pewny.

Ważną cechą pomeronu jest jego dodatnia parzystość ładunkowa.
Konsekwencją tego jest, że przy wysokich energiach, gdzie wnoszony przez nią wkład do amplitudy rozpraszania jest dominujący, amplituda rozproszenia $T_P$ jest taka sama w zderzeniach $pp$ jak w zderzeniach $p\bar p$.
Na pierwszy rzut oka może to wydawać się dziwne, ale należy pamiętać, że rozważania dotyczą jedynie oddziaływania silnego, więc znak ładunku elektrycznego nie gra tutaj roli.
Dodatkowo, wraz z~rosnącą energią struktura protonów i antyprotonów staje się zdominowana przez gluony, więc protony i antyprotony stają się coraz bardziej podobne do siebie.

I tutaj pojawia się \emph{odderon}.
Jest to hipotetyczna trajektoria podobna do pomeronu, ale o~ujemnej parzystości.
Konsekwencją jest to, że przyczynek $T_O$ wnoszonego przez wymianę odderonu do amplitud rozpraszania $pp$ i~$p\bar p$ ma przeciwny znak.
Warto nadmienić, że chromodynamika kwantowa sugeruje istnienie odderonu, ponieważ przewiduje ona, że gluony mogą tworzyć układy zarówno o parzystości dodatniej jak i ujemnej.

Uwzględniając zarówno wymianę pomeronu jak i odderonu, amplitudy oddziaływań $pp$ i $p\bar p$ wynoszą:
\begin{align}
T_N^{pp}     & = T_P + T_O\nonumber\\
T_N^{p\bar p} & = T_P - T_O
\end{align}
Istnienie odderonu oznacza, że oddziaływania $pp$ i $p \bar p$ będą się różnić nawet przy asymptotycznych energiach.

\section{Podsumowanie}

Rozpraszanie elastyczne jest kinematycznie najprostszym procesem zachodzącym w zderzeniach proton--proton,
jednak jego dynamika jest zaskakująco złożona. 
W oddziaływaniu silnym mamy do czynienia ze zjawiskami o naturze dyfrakcyjnej, co prowadzi do struktury minimum i maksimum obserwowanej w rozkładach kątowych.
Bardzo istotne jest twierdzenie optyczne wiążące amplitudę rozpraszania elastycznego.
Ponadto, dla bardzo dużych wartości $t$ oczekuje się przejścia w reżim perturbacyjny,
a dla małych wartości pojawiają się efekty oddziaływań elektromagnetycznych oraz zjawiska interferencji pomiędzy oddziaływaniem elektromagnetycznym i~silnym.

W pomiarach prowadzonych na akceleratorze LHC zastosowano dedykowany układ eksperymentalny ATLAS-ALFA.
Pomiary kątów rozpraszania rzędu pojedynczych mikroradianów wymagały zastosowania detektorów umieszczonych ponad 200 metrów od punktu zderzenia i zaledwie jeden milimetr od wiązki, co było możliwe dzięki technice rzymskich garnków.
Niezbędne też było użycie specjalnej optyki akceleratora LHC.

Wynikiem badań są podstawowe wielkości opisujące oddziaływania silne, w szczególności \stot i~$\rho$. 
Ich znajomość jest niezwykle istotna dla celów praktycznych, takich jak modelowanie zderzeń przy wysokich energiach.
Dostarczają one również wiedzy na temat oddziaływań silnych przy asymptotycznych energiach.
W szczególności mogą wskazywać na istnienie odderonu, niemniej tylko jako argument pośredni%
\footnote{Czytelników zainteresowanych tematem zachęcam do zapoznania się z artykułem \cite{D0TOTEM}, w którym zaprezentowano pomiary obszaru minimum i maksimum rozkładu $t$ w procesach elastycznych, w którym również można obserwować wpływ odderonu.}.

\section*{Podziękowania}

Bardzo dziękuję prof. Marioli Kłusek-Gawendzie, prof. Januszowi Chwastowskiemu oraz dr. Maciejowi Lewickiemu za cenne uwagi dotyczące artykułu.

Artykuł powstał w wyniku realizacji projektu badawczego SONATA BIS nr~2021/42/E/ST2/00350 finansowanego ze środków Narodowego Centrum Nauki.

\printbibliography[title={Literatura}]

@Article{ATLAS,
    author         = "{ATLAS Collaboration}",
    title          = "{The ATLAS Experiment at the CERN Large Hadron Collider}",
    journal        = "JINST",
    volume         = "3",
    year           = "2008",
    pages          = "S08003",
    doi            = "10.1088/1748-0221/3/08/S08003",
    primaryClass   = "hep-ex",
}

@Article{STDM-2018-08,
    author         = "{ATLAS Collaboration}",
    title          = "{Measurement of the total cross section and \(\rho\)-parameter from elastic scattering in \(pp\) collisions at \(\sqrt{s} = 13\,\text{TeV}\) with the ATLAS detector}",
    journal        = "Eur. Phys. J. C",
    volume         = "83",
    year           = "2023",
    pages          = "441",
    doi            = "10.1140/epjc/s10052-023-11436-8",
    reportNumber   = "CERN-EP-2022-129",
    eprint         = "2207.12246",
    archivePrefix  = "arXiv",
    primaryClass   = "hep-ex",
}

@article{ALFA,
    author = "Abdel Khalek, S. and others",
    title = "{The ALFA Roman Pot Detectors of ATLAS}",
    eprint = "1609.00249",
    archivePrefix = "arXiv",
    primaryClass = "physics.ins-det",
    doi = "10.1088/1748-0221/11/11/P11013",
    journal = "JINST",
    volume = "11",
    number = "11",
    pages = "P11013",
    year = "2016"
}

@Article{D0TOTEM,
    author = "{D0 and TOTEM Collaborations}",
    title = "{Odderon Exchange from Elastic Scattering Differences between $pp$ and $p \bar{p}$ Data at 1.96~TeV and from pp Forward Scattering Measurements}",
    eprint = "2012.03981",
    archivePrefix = "arXiv",
    primaryClass = "hep-ex",
    reportNumber = "FERMILAB-PUB-20-568-E, CERN-EP-2020-236",
    doi = "10.1103/PhysRevLett.127.062003",
    journal = "Phys. Rev. Lett.",
    volume = "127",
    number = "6",
    pages = "062003",
    year = "2021"
}

@article{COMPETE,
    author = "Cudell, J. R. and Ezhela, V. V. and Gauron, P. and Kang, K. and Kuyanov, Yu. V. and Lugovsky, S. B. and Martynov, E. and Nicolescu, B. and Razuvaev, E. A. and Tkachenko, N. P.",
    collaboration = "COMPETE",
    title = "{Benchmarks for the forward observables at RHIC, the Tevatron Run II and the LHC}",
    eprint = "hep-ph/0206172",
    archivePrefix = "arXiv",
    doi = "10.1103/PhysRevLett.89.201801",
    journal = "Phys. Rev. Lett.",
    volume = "89",
    pages = "201801",
    year = "2002"
}

\end{document}